# IMPACT OF COVID-19 ON THE BULLWHIP EFFECT ACROSS U.S. INDUSTRIES

**Alper Sarıcıoğlu [1,*], Müjde Erol Genevois [2], and Michele Cedolin [3]**

[1]Graduate School of Science and Engineering, Galatasaray University, 34349 Istanbul, Türkiye
[*]Corresponding author's e-mail: Alper Sarıcıoğlu (alper.saricioglu@gmail.com)

[2]Faculty of Engineering and Technology, Galatasaray University, 34349 Istanbul, Türkiye

[3]College of Engineering and Technology, American University of the Middle East, Egaila 54200, Kuwait

The Bullwhip Effect, describing the amplification of demand variability up the supply chain, poses significant challenges in Supply Chain Management. This study examines how the COVID-19 pandemic intensified the Bullwhip Effect across U.S. industries, using extensive industry-level data. By focusing on manufacturing, retailer, and wholesaler sectors, the research explores how external shocks exacerbate this phenomenon. Employing both traditional and advanced empirical techniques, the analysis reveals that COVID-19 significantly amplified the Bullwhip Effect, with industries displaying varied responses to the same external shock. These differences suggest that supply chain structures play a critical role in either mitigating or intensifying the effect. By analyzing the dynamics during the pandemic, this study provides valuable insights into managing supply chains under global disruptions and highlights the importance of tailoring strategies to industry-specific characteristics.

## 1. INTRODUCTION

The COVID-19 pandemic has significantly altered consumer behaviors, notably through panic buying and stockpiling. These reactions, influenced by perceived shortages and fears of ongoing crises, have led to substantial increases in demand for essential goods (Loxton *et al.* 2020, Yuen *et al.* 2020), and put considerable pressure on supply chains, particularly for essential products (Paul and Chowdhury 2021).

Simultaneously, the pandemic has highlighted supply chain vulnerabilities (Zhu *et al.* 2024), including the scarcity of input materials exacerbated by labor shortages due to health and safety restrictions (Maria del Rio-Chanona *et al.* 2020). Distinct industries, such as the automotive industry, have faced unique challenges due to their reliance on specific inputs (Ramani *et al.* 2022).

The logistical networks that support global supply chains also experienced severe disruptions, illustrating the fragility of these systems in the face of global crises (Xu *et al.* 2020). These issues were further compounded by international trade restrictions and transportation halts, leading to widespread economic impacts (Handfield *et al.* 2020).

Amid these demand shocks and supply disruptions, the Bullwhip Effect (BWE), a well-documented phenomenon in Supply Chain Management (SCM), describes the amplification of demand variability as orders move up the supply chain from end customers to suppliers was notably pronounced (Zighan 2022).

However, the relationship between COVID-19 and BWE is not limited to these drivers. The extraordinary circumstances of the pandemic significantly influenced the inventory decisions of buyers, as the heightened uncertainty made demand less predictable (Nikolopoulos *et al.* 2021). The BWE is inherently linked to the behaviors of supply chain planners, exemplified in the well-known beer game, a dynamic that illustrates how decision-makers overreact to changes in demand (Croson and Donohue 2006). Consequently, the difficulty in predicting future needs during the pandemic caused panic not only among consumers but also among inventory planners, who struggled to manage and respond to rapid shifts in demand (Gunessee and Subramanian 2020). Furthermore, the phenomenon of 'shortage gaming,' where industry players anticipate and react to the perceived scarcity of critical components, exacerbated supply shocks by artificially inflating demand, thus converting these supply-side disturbances into broader demand shocks (Seifert and Markoff 2020).

Traditionally associated with significant inefficiencies, including excess inventory, poor customer service levels, and lost revenues (Chen and Lee 2017), BWE has been the focus of numerous studies aiming to mitigate its impact under normal economic conditions (Brauch *et al.* 2024). However, the pandemic introduced unprecedented global disruptions, providing a novel context to explore the dynamics of the BWE under extreme external shocks (Queiroz *et al.* 2022).

While numerous empirical studies have explored the impact of COVID-19 on the economy (Remko 2020), the literature has yet to thoroughly examine how the BWE manifests across broader industry contexts during such pandemics. Therefore, there is a pressing need for empirical research that investigates the presence and magnitude of BWE during pandemics and dissects its manifestations across industries. Such studies would significantly enrich the literature, offering theoretical insights and practical guidance for managing supply chains in the face of global disruptions.

This study addresses this gap by providing detailed industry-level data for U.S. industries. By utilizing a comprehensive dataset encompassing all the U.S. industries, comprising manufacturing, wholesaler, and retailer stages, this research investigates the presence and magnitude of the BWE during the pandemic and dissects its manifestations across different sectors. At this point, industry-level data offer a unique vantage point, standing out for their broad scope and standardization, such as the North American Industry Classification System (NAICS).

The analysis employs a blend of traditional statistical methods and advanced machine learning (ML) techniques to forecast demand and inventory series, providing a robust framework for understanding the complex interplay between supply chain dynamics and external shocks. This approach allows for a nuanced exploration of industry-specific responses to the pandemic, contributing theoretical insights and practical guidance for managing supply chains in the face of global disruptions.

The study is structured as follows: Section 2 reviews relevant literature on the BWE and the impact of COVID-19 on supply chains, providing a theoretical backdrop for the analysis. Section 3 details the methodology, including the forecasting techniques and BWE analysis. Section 4 discusses the results, offering interpretations and implications for SCM. Finally, sections 5 and 6 summarize findings and suggest avenues for future research to investigate the resilience of supply chains to global disruptions.

## 2. LITERATURE REVIEW

### 2.1 The Bullwhip Effect: Demand Shocks and Supply Disruptions

The primary catalyst for the BWE under pandemic conditions is the inherent uncertainty in demand, which becomes particularly problematic in environments affected by frequent demand shocks. Such environments lead to less predictable demand patterns, which are challenging to forecast accurately, thus setting the stage for the amplification of order variability as inaccuracies cascade up the supply chain. Pastore et al. (2019) highlight that uncertainty in demand forecasts, often resulting from sudden demand shocks, substantially contributes to the BWE. The study by Springer and Kim (2010) specifically addresses how different inventory management policies react to sudden changes in demand and inadvertently exacerbate the BWE. Chen and Samroengraja (2004) study connect forecasting methods directly to the BWE, demonstrating how certain forecasting techniques in response to demand volatility can mitigate or worsen the effect.

As the pandemic has progressed, the reliability of traditional forecasting models has diminished, often leading supply chain managers to overreact to recent changes, thus amplifying demand variance. Traditional models, which rely heavily on historical data, have struggled to adapt to the rapid changes in market conditions brought about by the pandemic. Chiang et al. (2016) empirically analyzed the impacts of forecast accuracy, aggregate forecasting, and responsive forecasting on the BWE within the U.S. auto industry. Their findings indicate that while stable and reduced lead time forecasts can mitigate the effect, improvements in forecast accuracy do not necessarily result in reduced bullwhip impact. Similarly, Barlas and Gunduz (2011) demonstrated through system dynamics simulation that a major structural cause of the BWE is isolated demand forecasting performed at each supply chain echelon. According to Udenio et al. (2023), forecasting can cause the BWE primarily due to the inherent inaccuracies and the sensitivity of forecast parameters, especially under conditions of seasonal demand. Ma and Ma (2013) examine how different forecasting techniques affect the BWE in supply chains with two retailers. Their findings reveal that the choice of forecasting method substantially influences the BWE's magnitude. Jaipuria and Mahapatra (2014) explore advanced forecasting approaches combined with Artificial Neural Networks to mitigate the BWE in supply chains. Their study shows that this integration yields significantly more accurate demand forecasts than traditional ARIMA models, directly reducing the BWE by minimizing forecasting errors.

Ration gaming, a strategic behavior observed in supply chains, occurs when players artificially inflate their orders in anticipation of product shortages. This behavior is often triggered by the perception of limited supply, leading entities to order more than their actual demand to ensure sufficient product availability. Rong et al. (2008) investigate the causality of ration gaming in the context of BWE, utilizing a quantitative analysis in a supply chain setting. Their study demonstrates that retailers engage in ration gaming during upstream scarcity by increasing their order quantities beyond immediate needs in anticipation of future shortages. This behavior amplifies demand variability, BWE, as each retailer's inflated orders induce further over-ordering up the supply chain. Bray et al. (2019) explore the dynamics of ration gaming within a large retail supply chain comprising multiple stores and a single supplier. The study finds that ration gaming—where stores increase orders during anticipated shortages—directly contributes to the BWE.

Behavioral responses to supply chain conditions critically influence the severity of the BWE. Managers' reactions to perceived risks and ordering strategies in response to stock levels play pivotal roles in amplifying demand variability across the supply chain. Haines et al. (2017) delve into how awareness and control over one's cognitive processes can affect behaviors that lead to the BWE, particularly under the stress and uncertainty induced by the COVID-19 pandemic. Nienhaus et al. (2006) analyze the impact of human behavior on the BWE through a web-based simulation of the beer distribution. Their research highlights how human participants tend to amplify the BWE due to behavioral biases. Moritz et al. (2022) discuss how irrational ordering decisions, amplified during crises like the pandemic, lead to inefficient supply chain operations and heightened costs locally and across the entire supply chain network.

## 2.2 Empirical Research on BWE

The earliest empirical study that may contribute to BWE literature is rooted in the discussions of whether inventory has a smoothing or destabilizing effect. Holt et al. (1968), serially linked different stages of the television industry and measured fluctuations in material flow in terms of the mean absolute deviations. Blanchard (1983) compared the variance of production of U.S. car manufacturers and sales of their dealers and showed that the variance in production is larger than sales in both components. Blinder and Maccini (1991) suggested that the variance ratio between production and sales is always more than 1 for all stages: retailer, wholesaler, and manufacturing. Baganha and Cohen (1998) compared the coefficient of variation of each neighboring stage: production, manufacturing sales, wholesaler sales, and retailer sales in the U.S. Mollick (2004) using Japanese automotive industry data showed that nearly in all product segments, production variance was less than the sales variance. Beason (1993) reported that the sales variance was more than the production variance for nearly all 40 industries in Japan.

The first industry-level panel data study directly addressing BWE was introduced by Cachon et al. (2007). The results showed that 40% of manufacturing industries demonstrate BWE. Later Dooley et al. (2010) investigated how the 2007-2009 economic recession influenced inventory management practices and the BWE within the U.S. manufacturing sector. They utilized monthly inventory and sales data to examine how manufacturers, wholesalers, and retailers reacted to abrupt shifts in consumer demand.

The literature for empirical research for BWE can be categorized according to the data granularities (Yao *et al.* 2021). Single-firm data provide granular detail at the product and temporal levels; they lack the scope for assessing the phenomenon's prevalence across the economy (Chen and Lee 2012). Firm-level data, often derived from financial statements, offer firm-specific insights but fall short of temporal granularity and industry-wide representation (Shan *et al.* 2014, Bray and Mendelson 2015, Mackelprang and Malhotra 2015, Isaksson and Seifert 2016).

On the other hand, industry-level data allow for the examination of BWE across various sectors, presenting a more generalized view of its impact. This approach facilitates the analysis of industry-specific characteristics and their relation to BWE factors, leveraging a breadth of variables for a comprehensive analysis.

## 2.3 Panel Data Forecasting

Panel data forecasting is integral to preparing for and responding to economic disturbances. Integrating economic variables into forecasting models has provided valuable foresight during volatility (Fildes and Stekler 2002, Döpke *et al.* 2019). The seminal work by Baltagi (2008) provides a comprehensive foundation on the econometric models utilized in panel data analysis, highlighting the importance of capturing both cross-sectional and time-series dynamics in economic forecasting. Qu et al. (2023) discuss forecasting performance with panel data by comparing various econometric models. Their analysis suggests a persistent gap in capturing the temporal dynamics effectively when complex, nonlinear interactions exist between variables over time.

Contemporary methods, including ML and artificial intelligence applications in time series forecasting, have shown promising results in capturing complex, nonlinear economic relationships (Siami-Namini *et al.* 2018, Gökler and Boran 2022). Applying these new techniques, ranging from ARIMA to ML-based models such as LSTM networks, encapsulates a modern approach to tackling the variability inherent in SCM. ML approaches to economic forecasting have seen varied applications, as detailed by Gogas and Papadimitriou (2021) and Cicceri et al. (2020). They discuss the broader adoption of ML techniques in economics, including applications in recession forecasting and economic risk assessment. A recent study by Charles et al., (2023) contributes to the field by demonstrating the effectiveness of ML algorithms in accurately forecasting COVID-19 trends regarding confirmed, recovered, and death cases over six months. It highlights the robustness of these models across different geographical settings, providing valuable insights that can inform strategic decision-making and public health responses.

## 3. RESEARCH METHODOLOGY

This research is propelled by a dual motive: to seek empirical evidence of the BWE amidst the pandemic by examining industry-level panel data and to assess the ability of time series analysis to forecast the BWE in terms of amplification ratio.

Since BWE exists regardless of COVID-19, distinguishing between endogenous BWEs and those induced by external shocks of COVID-19 remains challenging. To tackle this, we employ advanced forecasting techniques to establish a baseline for the COVID-19 period, which represents what the scenario might have looked like without the pandemic's influence. This baseline is then compared with actual data to evaluate the vulnerability of specific industries to the BWE under pandemic conditions.

Figure 1 presents a comprehensive overview of this study's forecasting methodologies and BWE analysis.

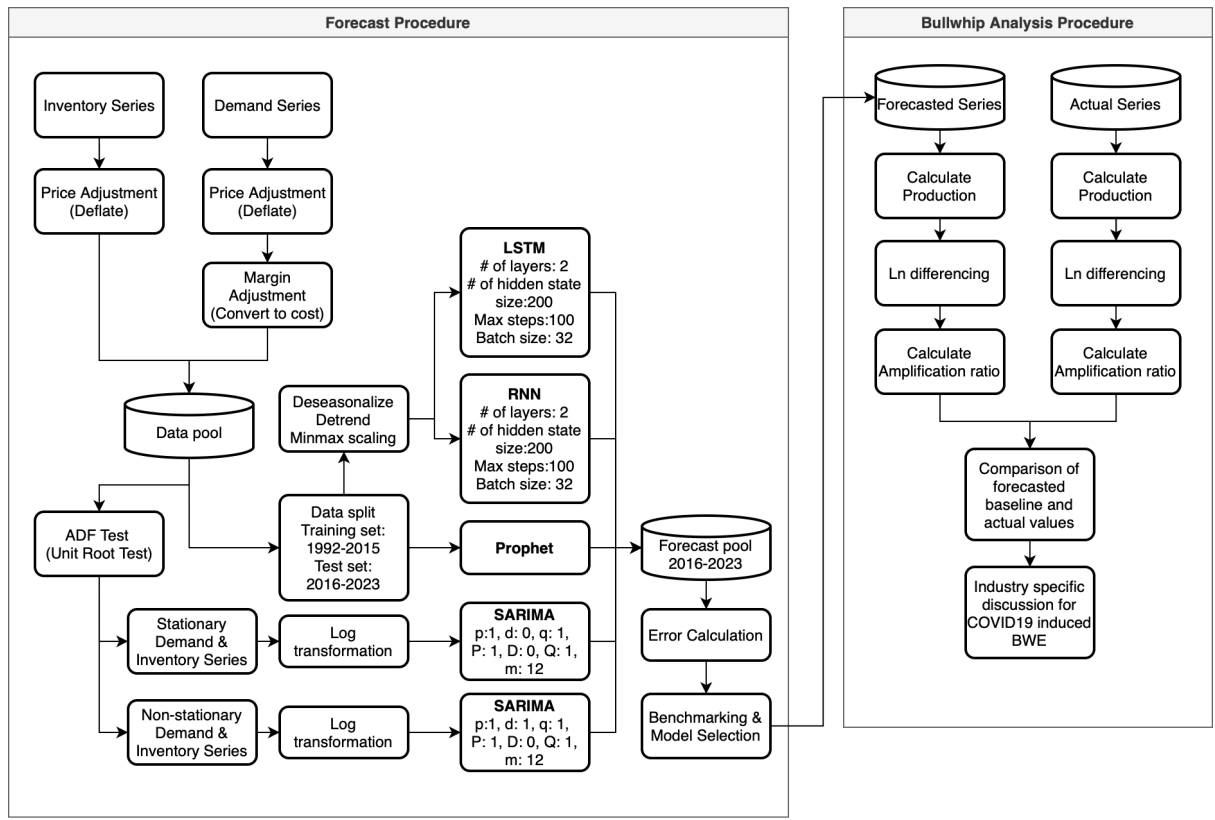

Figure 1 Flowchart of Procedures

The first stage involves preparing the raw inventory and demand data for analysis. Data was sourced from industry-level reports, including inventory and demand series. To ensure consistency, all data underwent price adjustments (deflation) using Implicit Price Deflators and margin adjustments to convert sales figures into cost values. Stationarity was assessed through the Augmented Dickey-Fuller (ADF) test, with stationary series proceeding directly to forecasting. Non-stationary series were log-transformed to stabilize variance and mean. Neural network models required additional preprocessing steps, including deseasonalization, detrending, and Minmax scaling, to ensure data compatibility. The preprocessed data was then pooled and split into training (1992–2015) and test (2016–2023) sets to evaluate model performance.

The forecasting phase involved generating demand and inventory forecasts for the test period (2016–2023) using preprocessed data. After splitting the data into training and test sets, models were trained on historical data (1992–2015) and tested for the COVID-19 period. Preprocessing steps, such as deseasonalization, detrending, and Minmax scaling for neural network models, ensured compatibility with their respective requirements. SARIMA, RNN, and LSTM models were tuned by specifying parameters such as autoregressive orders, hidden layers, and learning rates, as detailed in Figure 1. For Prophet, the default settings were used to allow the model to automatically identify changepoints, growth trends, and seasonalities. Forecasts from all models were aggregated into a forecast pool, and their accuracy was evaluated using metrics such as Mean Absolute Percentage Error (MAPE). The best-performing method was selected based on benchmarking results to ensure robust predictions for the subsequent Bullwhip Effect analysis.

In the third stage, BWE was analyzed by comparing the variance amplification between forecasted and actual series. Forecasted values served as a baseline to represent conditions without the impact of COVID-19, while actual values reflected the observed variances during the pandemic. Production levels were derived by calculating changes in inventory and demand. Log differencing was applied to the series to measure variance changes, and amplification ratios were calculated as the variance of production divided by the variance of demand. These ratios served as a quantitative measure of the BWE. The results were interpreted through a comparison of forecasted and actual amplification ratios across manufacturing, wholesale, and retail stages, highlighting the extent of the BWE and its variability across industries during the COVID-19 pandemic. The results were categorized into four zones by comparing the amplification ratio values in both the forecasted baseline and actual values. This categorization captured accurate and inaccurate predictions, providing insights into how the industries and supply chain stages were impacted by the pandemic. As the final step of the research, an industry-specific discussion for COVID-19-induced Bullwhip Effect (BWE) was conducted based on reported studies in the literature. This step analyzed how individual industries responded to pandemic-induced demand and supply shocks.

### 3.1 Data and Preprocessing

In this study, publicly available Manufacturers' Shipments, Inventories, & Orders (M3), Monthly Wholesale Trade (MWTS), and Monthly Retail Trade (MARTS) reports of U.S. Census Bureau are used. The dataset under examination provides detailed monthly demand and inventory values of the 77 U.S. industries from January 1992 to December 2023, from the manufacturing, retail, and wholesale stages. In addition, as a supporting dataset, Implicit Price Deflators for Manufacturing and Trade Sales of the Bureau of Economic Analysis are used to deflate the Demand and Inventory series.

The test data set is split into two periods: pre-COVID-19 (2016 – 2019) and COVID-19 (2020 – 2023). The aim was to identify the models that consistently deliver the most reliable and accurate prediction in the first period. Since the demand and supply shocks during the COVID-19 period can increase the errors as expected, the 2016-2019 period is considered a baseline to assess model performances.

The demand series has been price and margin-adjusted to translate sales figures into deflated cost values. For SARIMA, an Augmented Dickey-Fuller (ADF) test was carried out to determine the stationarity of the demand series. For non-stationary series, differencing and logarithmic conversions, are applied to stabilize variance and mean. Similar transformations are implemented for inventory series to ensure consistency in data. Since Neural Network based forecasting models perform poorly for seasonal and non-stationary demand (Zhang and Qi 2005), detrending and deseasonalizing are carried out as key steps in preparing time series data for RNN and LSTM methods along with Minmax normalization. Detrending removes long-term trends to stabilize the series' mean over time, which is important for models that need data to be stationary. The deseasonalizing step removed seasonal patterns, leaving behind the residual component that shows random variations. The original series is decomposed into an additive model's trend, seasonal, and residual components. After making predictions, the trend and seasonal components are added back to the predictions to make them interpretable and to ensure they reflect the original data's scale and behavior.

### 3.2 Forecasting Models

In the forecasting phase, Demand and Inventory series for 77 industries are forecasted with SARIMA, Prophet, RNN, and LSTM methods for COVID-19 impacted periods. The selection of the techniques in this study reflects an intentional effort to examine both statistical and machine learning (ML)-based forecasting approaches within the context of panel data and supply chain dynamics. SARIMA, a traditional statistical method, provides a reliable baseline for capturing seasonal and trend components. Prophet, while more automated, accommodates irregular events and multiple seasonality, making it suitable for diverse industry-level time series. In contrast, RNN and LSTM enable the exploration of non-linear dependencies and long-term patterns. By applying these techniques to panel data, this study assesses their performance in capturing both cross-sectional and temporal dynamics. This approach ensures a comprehensive comparison, leveraging traditional methods for foundational insights and ML-based methods for advanced, non-linear forecasting.

#### 3.2.1 SARIMA

SARIMA, is notable for its ability to explicitly incorporate seasonality through interpretable parameters, making it a benchmark for subsequent models. It is denoted as SARIMA*(p, d, q)(P, D,Q)m*, where *p* is the autoregressive, *d* is the differencing, and *q* is the moving average order, plus *(P,D,Q)* representing the seasonal autoregressive, seasonal differencing and seasonal moving average order, respectively. Finally, *m* is the number of periods in each season. The SARIMA equation is represented in Equation 1.

$$(1 - \sum_{i=1}^{p} \phi_i L^i)(1 - \sum_{i=1}^{P} \Phi_i L^{ixm})(1-L)^d (1-L^m)^D X_t = (1 + \sum_{i=1}^{q} \theta_i L^i)(1 + \sum_{i=1}^{Q} \Theta_i L^{ixm})\varepsilon_t \qquad (1)$$

where,
- $L$ is the lag operator, $L^k X_t = X_{t-k}$
- $\phi_i$, $\Phi_i$ are the coefficients for the autoregressive and seasonal autoregressive terms.
- $\theta_i$, $\Theta_i$ are the coefficients for the moving average and seasonal moving average terms.
- $X_t$ is the time series and $\varepsilon_t$ is the error term at time $t$

### 3.2.2 Prophet

Prophet is a forecasting tool developed by Facebook®, designed to produce high-quality forecasts for time series data that exhibit patterns on different time scales. It is particularly well-suited for data with strong seasonal effects and several seasons of historical data. In addition to SARIMA, Prophet can automatically detect trends, adjust the forecast to reflect growth that is not necessarily steady, and flexibly accommodate seasonality. The core formula of Prophet is given as Equation 2.

$$y_t = g_t + s_t + h_t + \epsilon_t \qquad (2)$$

where,
- $y_t$ is the predicted value at time $t$,
- $g_t$ is the trend component, modeling non-periodic changes over time,
- $s_t$ is the seasonality component, capturing periodic changes,
- $h_t$ is the effects of holidays or special events,
- $\epsilon_t$ is the error term at time $t$.

### 3.2.3 Recurrent Neural Network (RNN)

RNN, is an artificial neural network that recognizes patterns in data sequences. Unlike traditional neural networks that process inputs independently and assume inputs and outputs independence, RNNs have loops that allow information to persist, which is suited for tasks where context and temporal dynamics are crucial, like in time series forecasting. RNN involves an input layer where the data sequence is received, a hidden layer where the input is processed by combining it with the current state of the network, producing a new state, and an output layer that generates the next value of the time series. The fundamental formula for a basic RNN involves a simple update rule for the hidden state (Equation 3).

$$h_t = \sigma(W_h h_{t-1} + W_x x_t + b) \qquad (3)$$

where,
- $h_t$ is the hidden state at time step, *t* and $h_{t-1}$ is the hidden state at *t-1*
- $x_t$ is the input at *t*
- $W_h$ and $W_x$ are weights matrices for the hidden state and input, respectively.
- $b$ is the bias vector
- $\sigma$ is the activation function

### 3.2.4 Long Short-Term Memory (LSTM)

LSTM networks are a special kind of RNN model, capable of learning long-term dependencies in data sequences. Introduced by (Hochreiter and Schmidhuber 1997), LSTMs are specifically designed to avoid the long-term dependency problem typical of standard RNNs, making them highly effective for a wide range of sequence prediction problems where context from the distant past is informative for predicting future events. LSTM networks introduce a more complex update mechanism that involves several gates to control the memory and flow of information:

$$f_t = \sigma(W_f[h_{t-1}, x_t] + b_f) \tag{4}$$

$$i_t = \sigma(W_i[h_{t-1}, x_t] + b_i) \tag{5}$$

$$\tilde{C}_t = tanh(W_C[h_{t-1}, x_t] + b_C) \tag{6}$$

$$C_t = f_t(C_{t-1}) + i_t(\tilde{C}_t) \tag{7}$$

$$o_t = \sigma(W_o[h_{t-1}, x_t] + b_o) \tag{8}$$

$$h_t = o_t \tanh(C_t) \tag{9}$$

where,
- $f_t, i_t, o_t$ are forget, input, and output gates outputs in time step, *t*, respectively.
- $\tilde{C}_t$ is the new cell candidate vector,
- $C_t$ is the cell state vector,
- $h_t$ is the output hidden state vector,
- *W and b* represent weight matrices and bias vectors for different gates,
- *σ and tanh* are activation functions,

This study employs a systematic approach to forecasting demand and inventory series from 1992 to 2023. The methodologies summarized in Figure 1, integrated into our analysis, include traditional and advanced forecasting models. Monthly data from 1992 to 2016 was used to train forecasting models, and 2016 - 2023 data was used to evaluate their performance and confirm their predictions. The accuracy of forecasts is assessed using traditional non-scale-based error metrics, such as mean absolute percentage error (MAPE), which provides insights into each model's predictive accuracy and reliability, facilitating objective benchmarking and model selection process. Subsequently, for the Bullwhip Analysis, the amplification ratio for forecasted and actual values between 2020 and 2023 is calculated.

### 3.3 Bullwhip Effect Analysis Procedure

In harmony with the methodology of (Cachon *et al.* 2007), the production series, inferred from changes in inventory levels, are defined according to Equation 10 that relates production ($Y_{it}$), shipments ($S_{it}$), and inventory changes ($I_{it} - I_{it-1}$) over discrete time intervals *t* for industry *i*. This model, grounded in the principles outlined by (Blinder *et al.* 1981), posits that a decrease in inventory signals that the industry has engaged in production to meet demand, accounting for the net inventory reduction within the period. Conversely, an increase in inventory levels indicates that the industry has not only produced to meet demand but has also allocated additional resources to expand its inventory.

$$Y_{it} = S_{it} + (I_{it} - I_{it-1}) \tag{10}$$

The series are logged before the first differencing. The adjusted series is the approximate change percentage between observations in the series in Equation 11:

$$\ln(X_t) - \ln(X_{t-1}) \approx (X_t - X_{t-1})/X_{t-1} \tag{11}$$

where, *t* represents discrete time intervals, and $X_t$ denotes the time series variable, demand and production.

A common way to measure BWE is by calculating the ratio between the demand variances of upstream and downstream levels (Equation 12) (Disney and Towill 2003, Kim and Springer 2008, Cannella *et al.* 2013). The variances of demand and production were computed based on monthly values to calculate the amplification ratio.

$$Amplification\ ratio = \frac{Var_{Production}}{Var_{Demand}} \tag{12}$$

By the definition of the BWE, values greater than 1 represent BWE, and values between 0 and 1 mean that there is no BWE. Figure 2 shows how the amplification ratio is calculated by the demand and production of the same echelon. This method does not require a dyadic relationship between two parties. The amplification ratio can be measured individually regardless of whether it is a retailer (R), wholesaler (W), or manufacturer (M).

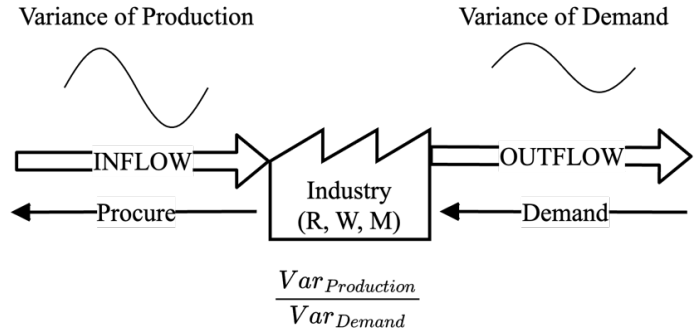

Figure 2 Measured Series

Since our data does not have a dyadic relationship between the two echelons, the production is assumed to be equivalent to its demand for upstream suppliers. In Equation 12, $Var_{Demand}$ defines the variance of demand series received by the industry and $Var_{Production}$ is production of the industry or demand of the industry for its suppliers.

## 4. RESULTS AND DISCUSSION

**4.1 Forecast Method Performance Comparison**

Table 1 shows the comparative analysis of forecasting models across two distinct periods, shown in 2016-2019 (pre-COVID-19) and 2020-2023 (during and post-COVID-19), respectively, with the average MAPE and MAPE variances of 77 industries both for Demand and Inventory series.

Table 1 Forecasting Results

|  | | 2016-2019 | | 2020-2023 | |
|---|---|---|---|---|---|
|  | Method | MAPE (%) | MAPE_Var (%) | MAPE (%) | MAPE_Var (%) |
| Demand | SARIMA | 9.05 | 0.74 | 21.46 | 3.78 |
|  | Prophet | 12.63 | 1.80 | 31.65 | 13.75 |
|  | RNN | 4.17 | 0.23 | 6.64 | 0.73 |
|  | LSTM | 4.07 | 0.22 | 6.63 | 0.72 |
| Inventory | SARIMA | 9.58 | 1.20 | 23.42 | 7.74 |
|  | Prophet | 12.64 | 1.68 | 31.78 | 10.86 |
|  | RNN | 1.86 | 0.02 | 2.60 | 0.03 |
|  | LSTM | 1.83 | 0.02 | 2.57 | 0.03 |

For Demand Forecasting, SARIMA and Prophet show significant increases in MAPE (21.46 % and 31.65%) during the COVID-19 period, indicating a decline in forecast accuracy under the unpredictable conditions brought about by the pandemic. The increase in MAPE_Var (3.78% and 13.75%) for these methods suggests that their forecasts became less consistent over time. In contrast, RNN and LSTM both demonstrate superior performance with much lower increases in MAPE from the pre-COVID-19 (4.17% and 4.07%) to the COVID-19 period (6.64% and 6.63%) compared to SARIMA and Prophet. They also maintain lower MAPE_Var values (0.73% and 0.72%), suggesting more stable forecasts despite the changing conditions.

Similar to demand forecasting, SARIMA, and Prophet exhibit larger errors and variances during COVID-19 (23.42% and 31.78%), reflecting challenges in adapting to the volatile environment. RNN and LSTM again outperform SARIMA and Prophet, maintaining remarkably low MAPE (2.60% and 2.57%) and MAPE_Var values even during the pandemic. This indicates high robustness and reliability in more complex and nonlinear scenarios.

Overall, RNN and LSTM show outstanding resilience to the disruptions caused by COVID-19, maintaining high accuracy and consistency in their forecasts for demand and inventory. Traditional models like SARIMA and Prophet struggled more with the volatility and irregularities brought about by the pandemic, leading to higher errors and inconsistencies. The data suggests that for handling more complex, nonlinear dynamics (especially in unpredictable scenarios like a pandemic), models like RNN and LSTM are preferable due to their ability to adapt and maintain performance.

Figure 3 and Figure 4 represent the MAPE distribution of all the forecasting models used in 77 industries for the Demand and Inventory Series, respectively.

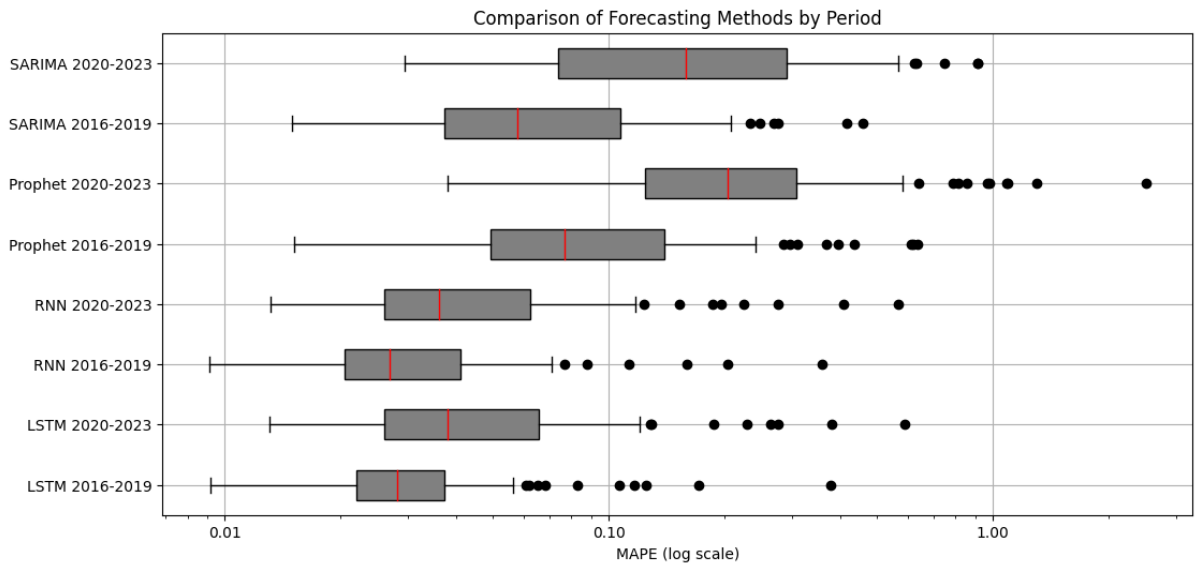

Figure 3 MAPE Boxplots (Demand)

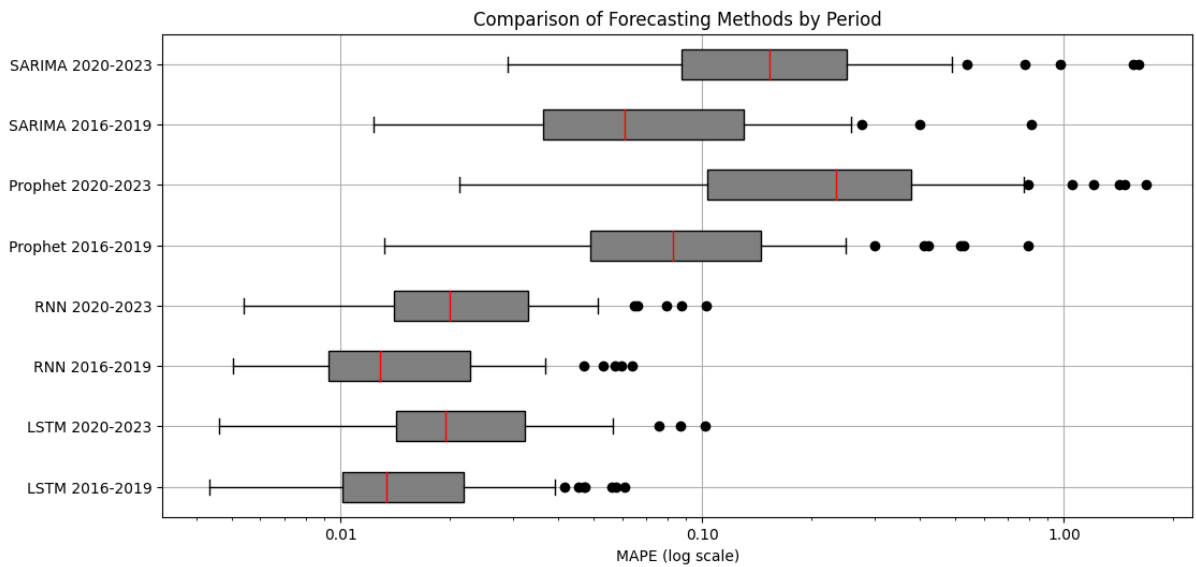

Figure 4 MAPE Boxplots (Inventory)

The performance of SARIMA significantly deteriorated in the 2020-2023 period, with a substantial increase in MAPE. RNN and LSTM showed consistent performance, although the error metrics slightly increased in the later period. This minimal increase suggests that these models maintained their predictive accuracy relatively well. The small rise in MAPE might be attributed to changes in underlying data patterns or external economic factors not captured in the model training phase. Prophet exhibited the most significant increase in error, with drastically higher MAPE values in the 2020-2023 period.

In conclusion, while RNN and LSTM stand out as superior forecasting tools for demand and inventory series, the LSTM forecasts edge slightly over RNN regarding consistency and reliability. Therefore, LSTM is selected as the baseline predictor for the Bullwhip Analysis stage.

**4.2 Amplification Ratios Comparison**

Figure 5 compares the actual amplification ratios against the predicted ones for 2020-2023, calculated with Equation 12. Whereas Actual Amplification Ratio is based on test data, LSTM Amplification Ratio is based on forecasted demand and inventory series.

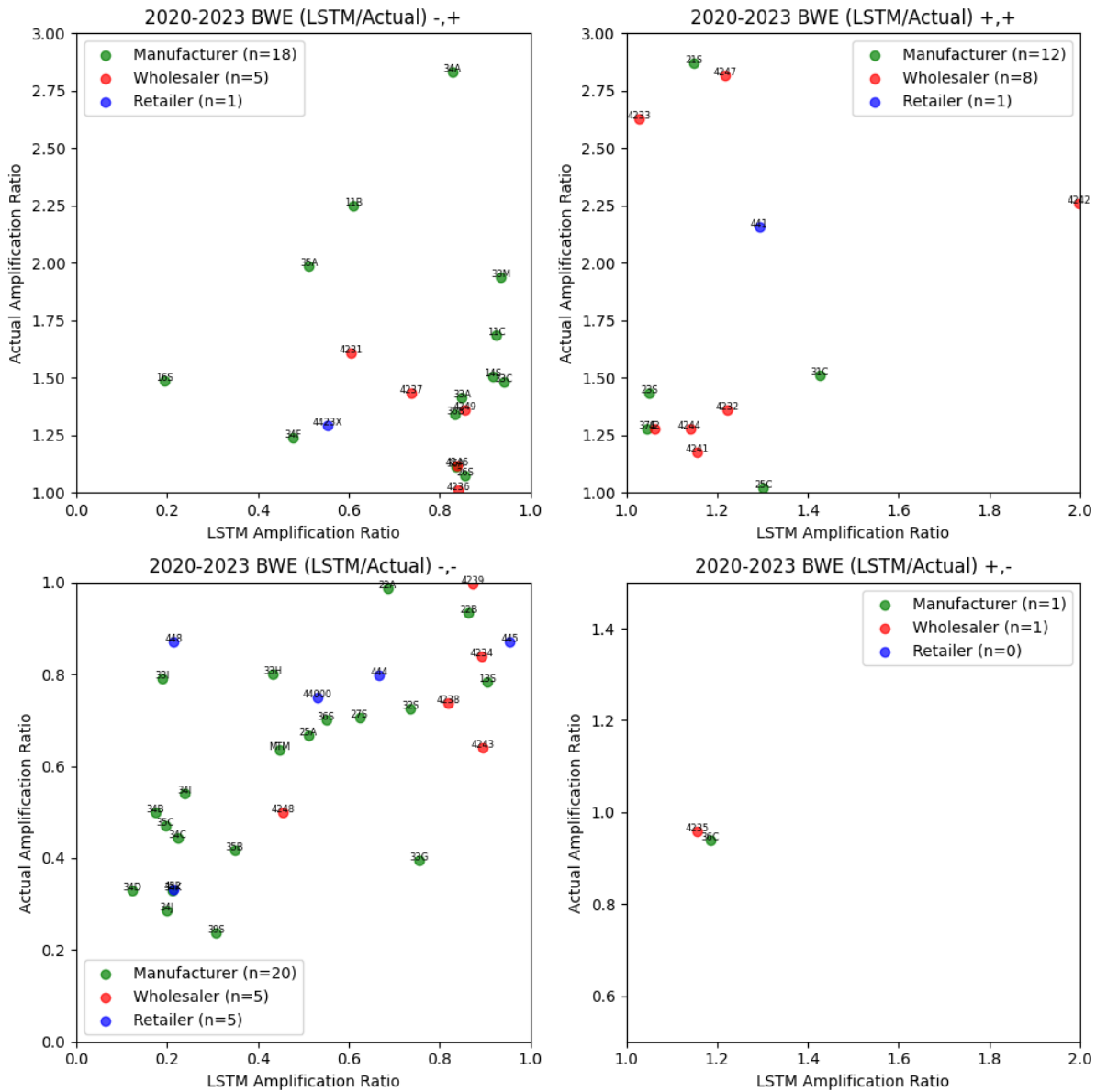

Figure 5 Amplification Ratios (Forecasted vs Actual)

The scatter diagram highlights the manufacturer, wholesaler, and retailer supply chain stages in different colors. In addition, each point in the scatter diagram is labeled with the industrial code. The diagram is divided into four zones, according to the calculated amplification ratios for the actual and predicted series, where values below 1 indicate no evidence for BWE (-), and values greater than 1 indicate BWE (+).

In the "Accurately Forecasted Bullwhip" zone (+,+), both forecasted and actual series indicate a BWE. "False Positive Bullwhip" (+,-) is populated with forecasts predicting a BWE, but the actual values did not substantiate this. Given that BWE is a distinct supply chain phenomenon, it should not be artificially generated by predictors. Nonetheless, forecasting errors can lead to inaccurately amplified variance. To ensure the robustness of a predictor, this category should ideally contain the fewest instances, as is observed in our study.

In the "Accurately Forecasted No Bullwhip" zone (-,-), both forecasted and actual values concur on the absence of a BWE, indicating that these industries are less frequently subjected to BWE systematically. "False Negative Bullwhip" (-,+), on the other hand, is comprised of forecasts that failed to predict a BWE that was indeed evident in the actual values. Industries within this category are a focal point of our analysis, particularly those potentially more impacted by COVID-19.

Table 2 summarizes the Figure 5 with aggregate results. The percentages for each stage are given next to the number of instances.

Table 2 LSTM Forecast and Actual BWE and Stages

| Stage | Accurately Forecasted Bullwhip (+,+) | Accurately Forecasted No Bullwhip (-,-) | False Positive Bullwhip (+,-) | False Negative Bullwhip (-,+) |
|---|---|---|---|---|
| Manufacturer | 12 (23.5%) | 20 (39.2%) | 1 (0.02%) | 18 (35.3%) |
| Retailer | 1 (14.2%) | 5 (71.4%) | - | 1 (14.2%) |
| Wholesaler | 8 (42.1%) | 5 (26.3%) | 1 (5.3%) | 5 (26.3%) |
| Total | 21 (27.3%) | 30 (39.0%) | 2 (2.6%) | 24 (31.2%) |

Manufacturers display a higher frequency of both Accurately Forecasted Bullwhip (12; 23.5%) False Negatives (-,+) (18; 35.3%) incidents, 58.8% of all manufacturing industries, suggesting a greater susceptibility to both the endogenous and external BWE. This aligns with the traditional understanding of the BWE, where demand variance amplification tends to propagate upstream, impacting manufacturers the most due to their position in the supply chain.

Wholesalers appear more prone to the BWE than retailers, as evidenced by their higher rates of False-negative incidents (5; 26.3%). This indicates that wholesalers are not immune to exogenous shocks, reflecting their intermediate position in the supply chain subjecting them to upstream supply variations and downstream demand fluctuations. However, showing moderate levels of both correct and missed predictions. Their intermediary position may offer better insights into supply and demand trends, helping mitigate some BWEs.

Retailers appear the least affected by the BWE with fewer instances (5; 71.4%) of the BWE, with a majority exhibiting variance smoothing. This observation is consistent with the BWE's theoretical framework, as retailers, being closer to the consumer end of the supply chain, typically experience less demand amplification.

The aggregated data reveal a general trend of higher bullwhip susceptibility upstream with manufacturers and a moderate impact on wholesalers. The lower impact on retailers highlights the effectiveness of proximity to the final consumer in mitigating the BWE. These insights emphasize the need for targeted strategies to address and mitigate the BWE, particularly for upstream supply chain actors.

## 4.3 Susceptibility of Industries Against Bullwhip Effect in COVID-19

The industries where the LSTM model did not predict the BWE (-,+) where it exists suggest a high susceptibility to this phenomenon, particularly during unpredictable pandemic conditions. Demand and supply shocks were identified using industry-specific reports on how sectors were affected by COVID-19, as referenced in the literature. These reports provided insights into changes in demand patterns, supply disruptions, and industry-specific dynamics during the pandemic. These industries experienced demand surges, demand stops, or supply disruptions due to labor or raw material shortages that can exacerbate the BWE, summarized in Table 3. The classification in Table 3 reflects these findings: upward arrows (↑) indicate increased demand or supply levels, downward arrows (↓) signify reduced demand or supply, and horizontal arrows (↔) represent industries with relatively stable demand or supply conditions.

Sample plots of demand series and corresponding predictions of forecast methods are given for each stage to illustrate the comparison between predicted and actual patterns.

Table 3 False Negative Bullwhip Industries

| Code | Industry | Demand Shock | Supply Shock |
|---|---|---|---|
| 11C | Meat, Poultry, and Seafood Product Processing (M) | Less affected ↔ | Negative ↓ |
| 26S | Plastics and Rubber Products (M) | | |
| 33A | Farm Machinery and Equipment Manufacturing (M) | | |
| 33E | Industrial Machinery Manufacturing (M) | | |
| 35A | Electric Lighting Equipment Manufacturing (M) | | |
| 35D | Battery Manufacturing (M) | | |
| 14S | Textile Products (M) | Negative ↓ | Negative ↓ |
| 15S | Apparel (M) | | |
| 16S | Leather and Allied Products (M) | | |
| 33C | Construction Machinery Manufacturing (M) | Negative ↓ | Less affected ↔ |
| 11B | Dairy Product Manufacturing (M) | Positive ↑ | Negative ↓ |
| 33M | Material Handling Equipment Manufacturing (M) | | |
| 34A | Electronic Computer Manufacturing (M) | | |
| 34F | Audio and Video Equipment (M) | | |
| 34H | Other Electronic Component Manufacturing (M) | | |
| 36A | Automobile Manufacturing (M) | | |
| 36B | Light Truck and Utility Vehicle Manufacturing (M) | | |
| 31A | Iron and Steel Mills and Ferroalloy and Steel Product (M) | Less affected ↔ | Negative ↓ |
| 4246 | Chemicals and Allied Products (W) | Less affected ↔ | Negative ↓ |
| 4237 | Hardware, and Plumbing and Heating Equipment and Supplies (W) | | |
| 4231 | Motor Vehicle and Motor Vehicle Parts and Supplies (W) | Negative ↓ | Negative ↓ |
| 4236 | Household Appliances and Electrical and Electronic Goods (W) | Positive ↑ | Negative ↓ |
| 4249 | Miscellaneous Nondurable Goods (W) | | |
| 4423X | Furniture, Home Furn., Electronics, and Appliance Stores (R) | Negative ↓ | Negative ↓ |

Eighteen manufacturer industries where the model did not predict a BWE that actually occurred are discussed below with the support of related literature. Demand for Meat, Poultry, and Seafood, Plastics and Rubber Products industries remained stable, driven by consistent consumer needs. However, the supply was challenged by outbreaks of COVID-19 in processing plants, leading to stringent health and safety measures to protect workers and maintain production levels (Brinca *et al.* 2021, Ramsey *et al.* 2021). Industrial Machinery, Farm Machinery and Equipment, Electric Lighting Equipment, and Battery Manufacturing industries were relatively unaffected as many sectors reliant on these industries and demand for agriculture continued during lockdowns, whereas supply suffered labor shortages (Maria del Rio-Chanona *et al.* 2020).

In the Textile Products, Apparel, and Leather and Allied Products Manufacturing industries, a sharp decline in demand was observed in fashion and luxury goods saw reduced consumer spending. On the supply side, production facilities faced operational restrictions due to disruptions caused by factory closures and restrictions in manufacturing hubs (Cho and Saki 2022, Seidu *et al.* 2023).

Construction Machinery Manufacturing industry experienced a decline in demand as non-essential construction projects were suspended in response to global lockdowns and social distancing measures (Brinca et al., 2021). These restrictions severely impacted the industry's ability to continue operations, leading to a drop in demand for construction machinery. Despite these challenges, the supply side was relatively more stable but not without disruptions, primarily due to reduced labor availability as workers were affected by health concerns and restrictions on movement (Ghansah and Lu 2023).

As (Liu and Rabinowitz 2021) study details, the imposition of stay-at-home orders and the closure of key institutional buyers like schools and restaurants shifted consumption patterns from dining out to home consumption, substantially increasing demand for Dairy Products. Demand for Material Handling Equipment and Light Truck and Utility Vehicle Manufacturing industries rose due to the expansion of e-commerce and increased automation needs in logistics operations (Herold *et al.* 2021). Audio and Video Equipment Manufacturing, Electronic Computer Manufacturing, and Other Electronic Component industries saw a surge in demand as consumers sought entertainment while confined to their homes, the surge in remote working and online schooling and eventually increasing need for electronic components in consumer electronics (Nikolopoulos *et al.* 2021). However, these industries faced supply disruptions due to the closure of manufacturing facilities, international logistics, and component shortages (Handfield *et al.* 2020, Walmsley *et al.* 2023).

Automobile Manufacturing initially experienced a dip in demand due to economic uncertainties (Figure 6). Demand subsequently increased as consumers began to prefer private vehicles over public transportation to minimize exposure to the virus (Basu and Ferreira 2021, Parker *et al.* 2021). The supply side in this sector was hampered by disruptions across the global supply chain and labor shortages (Belhadi *et al.* 2021).

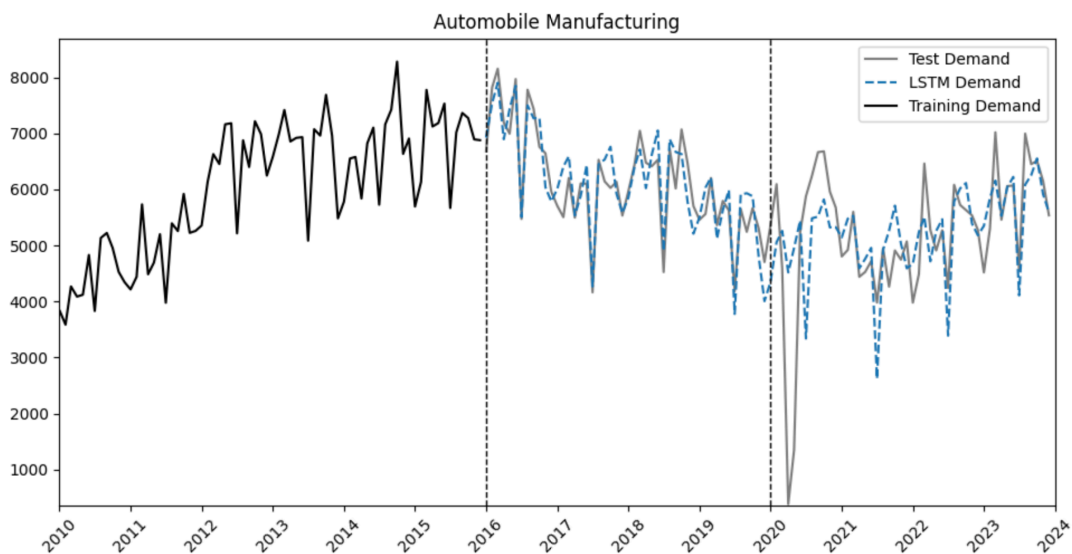

Figure 6 Demand Series for Automobile Manufacturing

While demand impacts were minimal, the Iron and Steel Mills and Ferroalloy and Steel Product Manufacturing industry struggled with supply chain disruptions from labor shortages, highlighting the need for more flexible production strategies (Asadi *et al.* 2023).

Five wholesaler industries where the BWE was observed, despite the did not predict it. Chemicals and Allied Products Wholesaling industry, critical in supplying essential materials across various sectors, faced unprecedented challenges due to global lockdown measures and border closures that disrupted supply chains and raw material availability. Demand patterns were notably altered. There was a surge in demand for chemicals necessary in pharmaceuticals and healthcare, including disinfectants and sanitizers (Abedsoltan 2023).

The Hardware, Plumbing, and Heating Equipment and Supplies industry exhibited trends similar to the Construction Machinery Manufacturing sector, where demand was negatively impacted due to the suspension of non-essential construction activities during lockdowns, while supply faced disruptions caused by component shortages (Brinca et al., 2021).

The Household Appliances and Electrical and Electronic Goods Wholesaling industry saw increased demand after an initial shock for home appliances due to more time spent at home, as shown in Figure 7. However, supply chains were challenged by manufacturing and international logistics disruptions, highlighting the need for supply chain visibility and flexibility (Nikolopoulos *et al.* 2021).

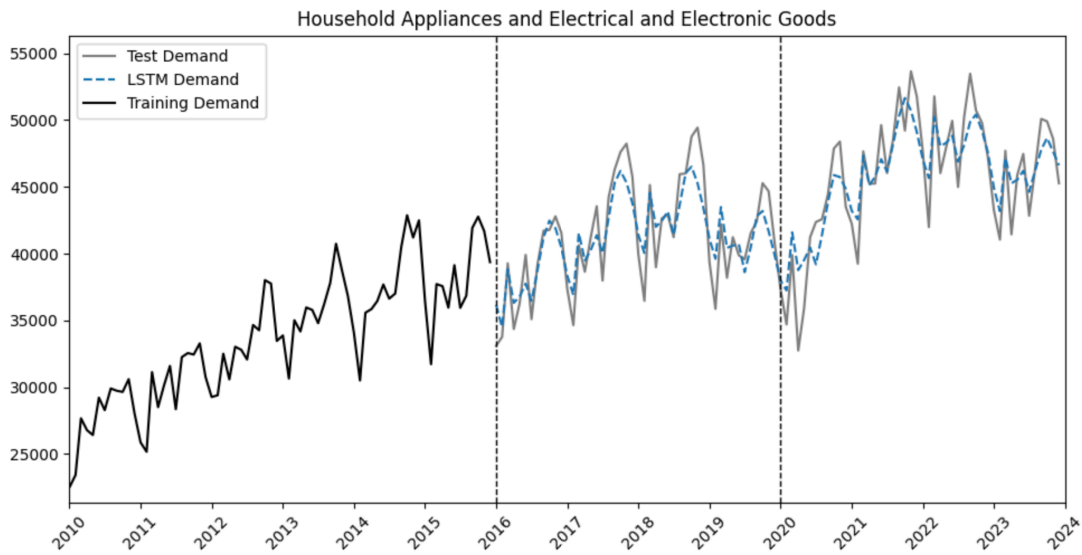

Figure 7 Demand Series for Household Appliances and Electrical and Electronic Goods Wholesaling

The Miscellaneous Nondurable Goods Wholesaling industry experienced demand spikes from panic buying, particularly for hygiene products, while supply was constrained by disruptions in manufacturing and logistics (Paul and Chowdhury 2020).

Motor Vehicle and Motor Vehicle Parts and Supplies Wholesaling suffered a demand decrease due to reduced car usage due to lockdowns. At the same time, supply issues stemmed from disruptions in global automotive manufacturing (Handfield *et al.* 2020).

Retailers appear the least affected by the BWE, which could be attributed to their end-of-chain position, allowing for more direct consumer demand assessments and rapid inventory and ordering strategy adjustments.

Furniture, Home Furnishings, Electronics, And Appliance Retailing industry experienced an initial demand shock in demand as consumers did not prioritize home improvement during lockdowns, as shown in Figure 8 (Verhoef *et al.* 2023). At the same time, the supply faced challenges due to the closures of retail outlets and disruptions in global manufacturing and shipping.

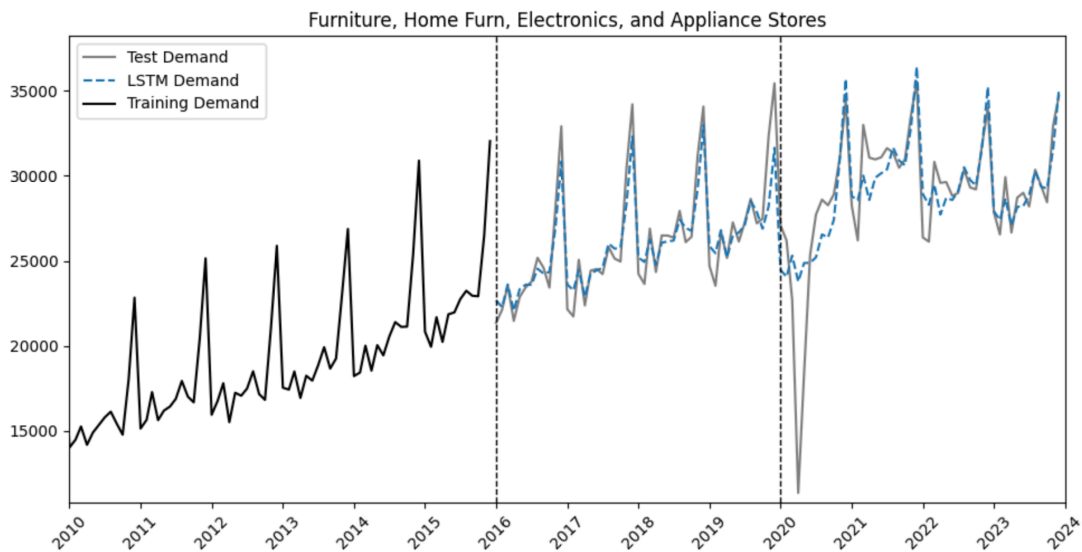

Figure 8 Demand Series for Furniture, Home Furnishings, Electronics, and Appliance Retailing

During the pandemic, diverse industries experienced significant shifts in demand and supply dynamics, reflecting the global crisis's broad impact and the varied nature of industrial operations.

The results suggest that the COVID-19 pandemic reshaped consumer behavior and demand across various sectors and highlighted vulnerabilities in global supply chains to trigger BWE. Supply disruptions have varied in nature and intensity across different industries, making it essential to study the BWE within a pandemic context to understand better how such external shocks can influence supply chain dynamics. In this context, the unprecedented disruptions brought forth by the pandemic provided a novel context for its examination, particularly how external shocks can exacerbate it.

## 5. CONCLUSION

This research has made significant strides in elucidating the dynamics of the BWE across U.S. industries during the tumultuous periods of the COVID-19 pandemic. It has been empirically demonstrated that the BWE, characterized by the amplification of demand variability up the supply chain, has indeed increased during these challenging times. Findings are crucial for understanding which industries were most affected, with a nuanced breakdown that maps these impacts within the context of demand and supply shocks. Furthermore, this study stands out as the first to employ industry-level data to forecast the BWE, revealing that it can be predicted during more stable periods, suggesting a new avenue for preemptively managing supply chain risks.

The study contributes to a more robust understanding of how global crises like the COVID-19 pandemic can exacerbate supply chain vulnerabilities. By revealing the specific industries that suffered the most, we provide insights that can help businesses and policymakers formulate strategies to bolster supply chain resilience against future disruptions. The predictive aspect of our research also introduces a proactive component to SCM, allowing stakeholders to anticipate and mitigate the impacts of the BWE before they reach critical levels.

## 6. FURTHER RESEARCH

The current study has highlighted that certain industries are more susceptible to the BWE during times of crisis. Future research should delve deeper into the specific characteristics contributing to this susceptibility. Researchers can provide more detailed insights into why certain industries face greater challenges by analyzing factors such as supply chain structure, dependency on critical components, inventory management strategies, and the flexibility of operational adjustments. This exploration could lead to a framework for assessing vulnerability across sectors, potentially guiding targeted strategies to bolster resilience.

While this research has utilized advanced forecasting models, the evolving field of data analytics and ML offers new tools that could enhance predictive accuracy and robustness. Future studies could explore integrating hybrid models that combine traditional econometric approaches with cutting-edge ML techniques, such as deep learning neural networks or ensemble methods that may improve forecast performance under various conditions. Comparing the effectiveness of these models in predicting the BWE could yield valuable methodologies for industry practitioners.

Further research should explore the long-term effects of the pandemic on supply chain dynamics and investigate additional strategies to enhance the robustness and adaptability of supply chains worldwide. This will not only help cushion against immediate shocks but also build enduring systems capable of thriving in an uncertain global economic landscape.

The focus on U.S. industries has provided detailed insights specific to the economic and supply chain dynamics within the U.S. Expanding this research to include industries in other countries or regions would enrich the understanding of the BWE under different economic, cultural, and regulatory environments. Such a comparative study could identify unique factors and commonalities affecting supply chain resilience globally, enabling a broader view of how global crises impact industries worldwide. Additionally, this approach might uncover strategies different nations employ to mitigate the effects of supply chain disruptions, offering a more comprehensive suite of solutions applicable on a global scale.

## 7. DISCLOSURE OF INTEREST

The authors confirm that there are no relevant financial or non-financial competing interests to report.
The authors declare that no funding was received for the conduct of this study and preparation of this article.

## ACKNOWLEDGMENT

This work was supported by Galatasaray University under Project FOA-2022-1128.

# 8. REFERENCES

Abedsoltan, H., 2023. COVID-19 and the chemical industry: impacts, challenges, and opportunities. *Journal of Chemical Technology and Biotechnology*, 98 (12).

Asadi, M., Tiwari, A.K., Gholami, S., Ghasemi, H.R., and Roubaud, D., 2023. Understanding interconnections among steel, coal, iron ore, and financial assets in the US and China using an advanced methodology. *International Review of Financial Analysis*, 89.

Baganha, M.P. and Cohen, M.A., 1998. The stabilizing effect of inventory in supply chains. *Operations Research*, 46 (3-supplement-3), S72–S83.

Baltagi, B.H., 2008. Forecasting with panel data. *Journal of Forecasting*, 27 (2).

Barlas, Y. and Gunduz, B., 2011. Demand forecasting and sharing strategies to reduce fluctuations and the bullwhip effect in supply chains. *Journal of the Operational Research Society*, 62 (3).

Basu, R. and Ferreira, J., 2021. Sustainable mobility in auto-dominated Metro Boston: Challenges and opportunities post-COVID-19. *Transport Policy*, 103.

Beason, R., 1993. Tests of production smoothing in selected Japanese industries. *Journal of Monetary Economics*, 31 (3), 381–394.

Belhadi, A., Kamble, S., Jabbour, C.J.C., Gunasekaran, A., Ndubisi, N.O., and Venkatesh, M., 2021. Manufacturing and service supply chain resilience to the COVID-19 outbreak: Lessons learned from the automobile and airline industries. *Technological Forecasting and Social Change*, 163.

Blanchard, O.J., 1983. The Production and Inventory Behavior of the American Automobile Industry. *Journal of Political Economy*, 91 (3), 365–400.

Blinder, A.S., Lovell, M.C., and Summers, L.H., 1981. Retail Inventory Behavior and Business Fluctuations. *Brookings Papers on Economic Activity*, 1981 (2), 443–520.

Blinder, A.S. and Maccini, L.J., 1991. Taking Stock: A Critical Assessment of Recent Research on Inventories. *Journal of Economic Perspectives*, 5 (1), 73–96.

Brauch, M., Mohaghegh, M., and Größler, A., 2024. Causes of the bullwhip effect: a systematic review and categorization of its causes. *Management Research Review*.

Bray, R.L. and Mendelson, H., 2015. Production smoothing and the bullwhip effect. *Manufacturing and Service Operations Management*, 17 (2), 208–220.

Bray, R.L., Yao, Y., Duan, Y., and Huo, J., 2019. Ration gaming and the bullwhip effect. *Operations Research*, 67 (2).

Brinca, P., Duarte, J.B., and Faria-e-Castro, M., 2021. Measuring labor supply and demand shocks during COVID-19. *European Economic Review*, 139.

Cachon, G.P., Randall, T., and Schmidt, G.M., 2007. In search of the bullwhip effect. *Manufacturing and Service Operations Management*, 9 (4), 457–479.

Cannella, S., Barbosa-Póvoa, A.P., Framinan, J.M., and Relvas, S., 2013. Metrics for bullwhip effect analysis. *Journal of the Operational Research Society*, 64 (1).

Charles, V., Mousavi, S.M.H., Gherman, T., and Mosavi, S.M.H., 2023. From data to action: Empowering COVID-19 monitoring and forecasting with intelligent algorithms. *Journal of the Operational Research Society*.

Chen, F. and Samroengraja, R., 2004. Order volatility and supply chain costs. *Operations Research*, 52 (5).

Chen, L. and Lee, H.L., 2012. Bullwhip effect measurement and its implications. *Operations Research*, 60 (4), 771–784.

Chen, L. and Lee, H.L., 2017. Modeling and Measuring the Bullwhip Effect. *In*: *Springer Series in Supply Chain Management*.

Chiang, C.Y., Lin, W.T., and Suresh, N.C., 2016. An empirically-simulated investigation of the impact of demand forecasting on the bullwhip effect: Evidence from U.S. auto industry. *International Journal of Production Economics*, 177, 53–65.

Cho, B. and Saki, Z., 2022. Firm performance under the COVID-19 pandemic: The case of the U.S. textile and apparel industry. *Journal of the Textile Institute*, 113 (8).

Cicceri, G., Inserra, G., and Limosani, M., 2020. A machine learning approach to forecast economic recessions-an Italian case study. *Mathematics*, 8 (2).

Croson, R. and Donohue, K., 2006. Behavioral Causes of the Bullwhip Effect and the Observed Value of Inventory Information. *Management Science*, 52 (3), 323–336.

Disney, S.M. and Towill, D.R., 2003. On the bullwhip and inventory variance produced by an ordering policy. *Omega*, 31 (3).

Dooley, K.J., Yan, T., Mohan, S., and Gopalakrishnan, M., 2010. Inventory management and the bullwhip effect during the 2007-2009 recession: Evidence from the manufacturing sector. *Journal of Supply Chain Management*.

Döpke, J., Fritsche, U., and Müller, K., 2019. Has macroeconomic forecasting changed after the Great Recession? Panel-based evidence on forecast accuracy and forecaster behavior from Germany. *Journal of Macroeconomics*, 62.

Fildes, R. and Stekler, H., 2002. The state of macroeconomic forecasting. *Journal of Macroeconomics*, 24 (4).

Ghansah, F.A. and Lu, W., 2023. Responses to the COVID-19 pandemic in the construction industry: a literature review of academic research. *Construction Management and Economics*, 41 (9).

Gogas, P. and Papadimitriou, T., 2021. Machine Learning in Economics and Finance. *Computational Economics*.

Gökler, S.H. and Boran, S., 2022. Prediction of Demand for Red Blood Cells Using Ridge Regression, Artificial Neural Network, and Integrated Taguchi-Artificial Neural Network Approach. *International Journal of Industrial Engineering : Theory Applications and Practice*, 29 (1).

Gunessee, S. and Subramanian, N., 2020. Ambiguity and its coping mechanisms in supply chains lessons from the Covid-19 pandemic and natural disasters. *International Journal of Operations and Production Management*, 40 (7–8).

Haines, R., Hough, J., and Haines, D., 2017. A metacognitive perspective on decision making in supply chains: Revisiting the behavioral causes of the bullwhip effect. *International Journal of Production Economics*, 184.

Handfield, R.B., Graham, G., and Burns, L., 2020. Corona virus, tariffs, trade wars and supply chain evolutionary design. *International Journal of Operations and Production Management*, 40 (10).

Herold, D.M., Nowicka, K., Pluta-Zaremba, A., and Kummer, S., 2021. COVID-19 and the pursuit of supply chain resilience: reactions and "lessons learned" from logistics service providers (LSPs). *Supply Chain Management*, 26 (6).

Hochreiter, S. and Schmidhuber, J., 1997. Long Short Term Memory. *Neural Computation*, 9 (8).

Holt, C.C., Modigliani, F., and Shelton, J.P., 1968. The Transmission of Demand Fluctuations Through a Distribution and Production System, the Tv-Set Industry. *The Canadian Journal of Economics*, 1 (4), 718.

Isaksson, O.H.D. and Seifert, R.W., 2016. Int . J . Production Economics Quantifying the bullwhip effect using two-echelon data : A cross-industry empirical investigation. *Intern. Journal of Production Economics*, 171, 311–320.

Jaipuria, S. and Mahapatra, S.S., 2014. An improved demand forecasting method to reduce bullwhip effect in supply chains. *Expert Systems with Applications*, 41 (5).

Kim, I. and Springer, M., 2008. Measuring endogenous supply chain volatility: Beyond the bullwhip effect. *European Journal of Operational Research*, 189 (1).

Liu, Y. and Rabinowitz, A.N., 2021. The impact of the COVID-19 pandemic on retail dairy prices. *Agribusiness*, 37 (1).

Loxton, M., Truskett, R., Scarf, B., Sindone, L., Baldry, G., and Zhao, Y., 2020. Consumer Behaviour during Crises: Preliminary Research on How Coronavirus Has Manifested Consumer Panic Buying, Herd Mentality, Changing Discretionary Spending and the Role of the Media in Influencing Behaviour. *Journal of Risk and Financial Management*, 13 (8).

Ma, J. and Ma, X., 2013. A comparison of bullwhip effect under various forecasting techniques in supply chains with two retailers. *Abstract and Applied Analysis*, 2013.

Mackelprang, A.W. and Malhotra, M.K., 2015. The impact of bullwhip on supply chains: Performance pathways, control mechanisms, and managerial levers. *Journal of Operations Management*, 36, 15–32.

Maria del Rio-Chanona, R., Mealy, P., Pichler, A., Lafond, F., and Doyne Farmer, J., 2020. Supply and demand shocks in the COVID-19 pandemic: An industry and occupation perspective. *Oxford Review of Economic Policy*.

Mollick, A.V., 2004. Production smoothing in the Japanese vehicle industry. *International Journal of Production Economics*, 91 (1), 63–74.

Moritz, B.B., Narayanan, A., and Parker, C., 2022. Unraveling Behavioral Ordering: Relative Costs and the Bullwhip Effect. *Manufacturing and Service Operations Management*, 24 (3).

Nienhaus, J., Ziegenbein, A., and Schoensleben, P., 2006. How human behaviour amplifies the bullwhip effect. A study based on the beer distribution game online. *Production Planning and Control*, 17 (6), 547–557.

Nikolopoulos, K., Punia, S., Schäfers, A., Tsinopoulos, C., and Vasilakis, C., 2021. Forecasting and planning during a pandemic: COVID-19 growth rates, supply chain disruptions, and governmental decisions. *European Journal of Operational Research*, 290 (1).

Parker, M.E.G., Li, M., Bouzaghrane, M.A., Obeid, H., Hayes, D., Frick, K.T., Rodríguez, D.A., Sengupta, R., Walker, J., and Chatman, D.G., 2021. Public transit use in the United States in the era of COVID-19: Transit riders' travel behavior in the COVID-19 impact and recovery period. *Transport Policy*, 111.

Pastore, E., Alfieri, A., and Zotteri, G., 2019. *An empirical investigation on the antecedents of the bullwhip effect: Evidence from the spare parts industry*. International Journal of Production Economics. Elsevier B.V.

Paul, S.K. and Chowdhury, P., 2020. Strategies for Managing the Impacts of Disruptions During COVID-19: an Example of Toilet Paper. *Global Journal of Flexible Systems Management*, 21 (3).

Paul, S.K. and Chowdhury, P., 2021. A production recovery plan in manufacturing supply chains for a high-demand item during COVID-19. *International Journal of Physical Distribution and Logistics Management*, 51 (2).

Qu, R., Timmermann, A., and Zhu, Y., 2023. Comparing forecasting performance with panel data. *International Journal of Forecasting*.

Queiroz, M.M., Ivanov, D., Dolgui, A., and Fosso Wamba, S., 2022. Impacts of epidemic outbreaks on supply chains: mapping a research agenda amid the COVID-19 pandemic through a structured literature review. *Annals of Operations Research*, 319 (1).

Ramani, V., Ghosh, D., and Sodhi, M.M.S., 2022. Understanding systemic disruption from the Covid-19-induced semiconductor shortage for the auto industry. *Omega (United Kingdom)*, 113.

Ramsey, A.F., Goodwin, B.K., Hahn, W.F., and Holt, M.T., 2021. Impacts of COVID-19 and Price Transmission in U.S. Meat Markets. *Agricultural Economics (United Kingdom)*, 52 (3).

Remko, van H., 2020. Research opportunities for a more resilient post-COVID-19 supply chain – closing the gap between research findings and industry practice. *International Journal of Operations and Production Management*, 40 (4).

Rong, Y., Snyder, L. V., and Shen, Z.-J.M., 2008. Bullwhip and Reverse Bullwhip Effects under the Rationing Game. *SSRN Electronic Journal*.

Seidu, R.K., Jiang, S. xiang, Tawiah, B., Acquaye, R., and Howard, E.K., 2023. Review of effects of COVID-19 pandemic on the textile industry: challenges, material innovation and performance. *Research Journal of Textile and Apparel*.

Seifert, R.W. and Markoff, R., 2020. Digesting the shocks: how supply chains are adapting to the COVID-19 lockdowns. *Imd*.

Shan, J., Yang, S.S., Yang, S.S., and Zhang, J., 2014. An empirical study of the bullwhip effect in China. *Production and Operations Management*, 23 (4), 537–551.

Siami-Namini, S., Tavakoli, N., and Siami Namin, A., 2018. A Comparison of ARIMA and LSTM in Forecasting Time Series. *In*: *Proceedings - 17th IEEE International Conference on Machine Learning and Applications, ICMLA 2018*.

Springer, M. and Kim, I., 2010. Managing the order pipeline to reduce supply chain volatility. *European Journal of Operational Research*, 203 (2).

Udenio, M., Vatamidou, E., and Fransoo, J.C., 2023. Exponential smoothing forecasts: taming the bullwhip effect when demand is seasonal. *International Journal of Production Research*, 61 (6).

U.S. Census Bureau, 2024. Business & Industry Time Series / Trend Charts [online]. Available from: https://www.census.gov/econ_timeseries/?programCode=RESSALES&startYear=1963&endYear=2024&categories[]=ASOLD&dataType=TOTAL&geoLevel=US&adjusted=1¬Adjusted=0&errorData=0 [Accessed 2 May 2024].

Verhoef, P.C., Noordhoff, C.S., and Sloot, L., 2023. Reflections and predictions on effects of COVID-19 pandemic on retailing. *Journal of Service Management*, 34 (2).

Walmsley, T., Rose, A., John, R., Wei, D., Hlávka, J.P., Machado, J., and Byrd, K., 2023. Macroeconomic consequences of the COVID-19 pandemic. *Economic Modelling*, 120.

Xu, Z., Elomri, A., Kerbache, L., and El Omri, A., 2020. Impacts of COVID-19 on Global Supply Chains: Facts and Perspectives. *IEEE Engineering Management Review*, 48 (3).

Yao, Y., Duan, Y., and Huo, J., 2021. On empirically estimating bullwhip effects: Measurement, aggregation, and impact. *Journal of Operations Management*, 67 (1), 5–30.

Yuen, K.F., Wang, X., Ma, F., and Li, K.X., 2020. The psychological causes of panic buying following a health crisis. *International Journal of Environmental Research and Public Health*.

Zhang, G.P. and Qi, M., 2005. Neural network forecasting for seasonal and trend time series. *In*: *European Journal of Operational Research*.

Zhu, X., Wen, Z., Regan, D., and Zhu, J., 2024. A Systematic Analysis of Supply Chain Risk Management Literature: 2012-2021. *International Journal of Industrial Engineering: Theory, Applications and Practice*, 31 (3).

Zighan, S., 2022. Managing the great bullwhip effects caused by COVID-19. *Journal of Global Operations and Strategic Sourcing*, 15 (1).